\documentclass[conference]{IEEEtran}
\IEEEoverridecommandlockouts
\usepackage{cite}
\usepackage{amsmath,amssymb,amsfonts}
\usepackage{algorithmic}
\usepackage{graphicx}
\usepackage{textcomp}
\usepackage{xcolor}
\usepackage{enumitem,kantlipsum}

\def\BibTeX{{\rm B\kern-.05em{\sc i\kern-.025em b}\kern-.08em
    T\kern-.1667em\lower.7ex\hbox{E}\kern-.125emX}}
    
\begin{document}
	\IEEEspecialpapernotice{(Invited Paper)}
\title{Physical Channel Modeling for RIS-Empowered Wireless Networks in Sub-6 GHz Bands}

\author{\IEEEauthorblockN{Fatih Kilinc\textsuperscript{$\ast$}, Ibrahim Yildirim\textsuperscript{$\ast$,$\bullet$}, and Ertugrul Basar\textsuperscript{$\ast$}}
\IEEEauthorblockA{\textsuperscript{$\ast$}CoreLab, Department of Electrical and Electronics Engineering, Koç University, Sariyer 34450, Istanbul, Turkey \\
\textsuperscript{$\bullet$}Faculty of Electrical and Electronics Engineering, Istanbul Technical University, Sariyer 34469, Istanbul, Turkey.
\\ Email: fkilinc20@ku.edu.tr, yildirimib@itu.edu.tr, ebasar@ku.edu.tr}
}

\maketitle

\begin{abstract}
Reconfigurable intelligent surface (RIS)-assisted communications is one of the promising candidates for next generation wireless networks by controlling the propagation environment dynamically. In this study, a channel modeling strategy for RIS-assisted wireless networks is introduced in sub-6 GHz bands by considering both far-field and near-field behaviours in transmission. We also proposed an open-source physical channel simulator for sub-6 GHz bands where operating frequency, propagation environment, terminal locations, RIS location and size can be adjusted. It is demonstrated via extensive computer simulations that an improved achievable rate performance is obtained in the presence of RISs for both near-field and far-field conditions.
\end{abstract}

\section{Introduction}
Due to the developments in technology and the increase in data consumption, current communication systems have been inadequate to meet the increasing demands. Hence, next-generation wireless systems should provide data transfer at very high rates with a high reliability. Emerging technologies such as virtual-reality, high-quality streaming, and autonomous systems require novel ways to achieve higher data rates with low latency. Massive multiple-input multiple-output (mMIMO), millimeter-wave (mmWave) communications, terahertz communications, and reconfigurable intelligent surfaces (RISs) can be considered as remarkable technologies for next-generation communication systems to enhance system performance\cite{basar2019wireless}.

RISs have recently received great attention due to their advantages in controlling the wireless propagation environment. An RIS is comprised of small electro-magnetic (EM) surfaces, which are capable of reflecting the incoming signals in a controllable and efficient way, and provides a more energy-efficient wireless communication network. The reflection properties of the EM surfaces can be dynamically controlled with nearly passive electronic components, called "meta-atoms".
Considering all of these advantages of RIS, the implementation of an RIS in mmWave bands is one of the cost and power-efficient ways to enhance the system performance. 

RISs have been extensively investigated in the literature from various aspects, where the physical properties, behaviours and different use cases of RISs, such as holographic beamformers, low complexity transmitters, anomalous reflectors are investigated \cite{basar2019wireless,huang2020holographic}. The integration of an RIS into currently existing technologies such as index modulation (IM) techniques and relay-aided systems are proposed in \cite{yildirim2021hybrid} and \cite{basar2020reconfigurable}. Moreover, the efficient positioning and beamforming optimization of the RIS in MIMO networks are studied in \cite{wu2020intelligent} and \cite{zhang2020capacity}.
Since carrier signals are vulnerable to blockage and suffer from path loss in mmWave communications, an RIS can be used to ensure reliable transmission by creating an additional path. Therefore, the most of the studies in the literature are focused on the deployment of RIS in mmWave frequencies.  Modeling of the RIS-assisted wireless channels also stands out as another key direction for the research community. RISs are mostly designed in the form of uniform planar array (UPA), therefore, two dimensional (2D) channel models are not accurate for RIS-assisted networks. In this context, 3D physical channel models for RIS-assisted mmWave networks are proposed in several recent works \cite{basar2020indoor,basar2020simris,yildirim2020modeling}. However, the RIS technology can also be used in sub-6 GHz bands, which is used for majority of wireless communication systems, such as 3G/4G and even 5G cellular networks, Wi-Fi systems in our homes, and bluetooth devices. The existing studies, which reveal the effects and behavior of RIS in sub-6 GHz bands, have not exhaustively investigated physical channels. 

In this study, we propose a modeling strategy for RIS-empowered communication systems by considering the currently used technical specifications on sub-6 GHz bands \cite{3GPP_Sub6G,article}. According to a common assumption in the literature, the most efficient use of an RIS is possible when it is placed close to the terminals. Since the wavelengths of the carriers in sub-6 GHz bands are significantly higher than those in mmWave bands, deployment of a large RIS in sub-6 GHz bands requires the RIS to operate in near-field if the RIS is placed close to transmitter (Tx) or  receiver (Rx). In order to cover these use-cases, the impact of the RIS in both far-field and near-field conditions is analyzed in this study.

The rest of this paper can be summarized as follows. In Section II, we propose RIS-assisted end-to-end channel model for indoor and outdoor environments operating in sub-6 GHz bands. In Section III, we provide the achievable rate analysis  of the system. In Section IV, the numerical results are presented and the paper is concluded in Section V.

\begin{figure*}[!t]
	\begin{center}
		\includegraphics[width=1.9\columnwidth]{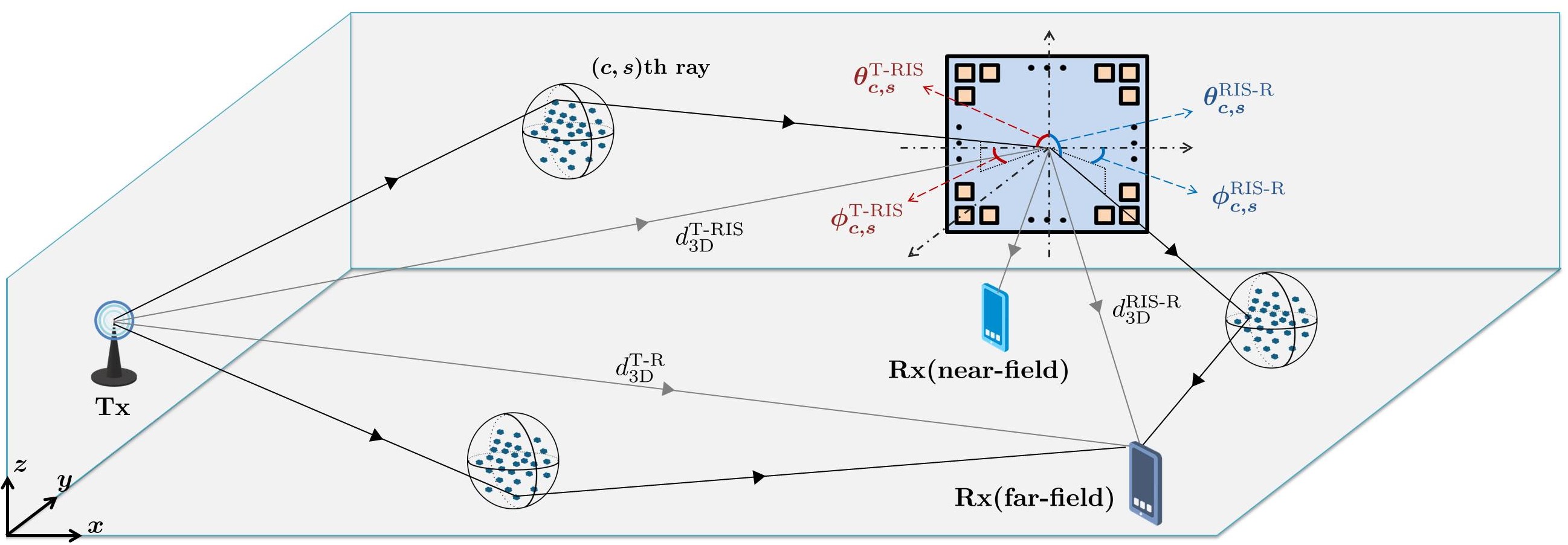}
		\vspace*{-0.3cm}\caption{Generic system model for an RIS-assisted network with $C$ number of clusters and $S$ number of rays.}\vspace*{-0.3cm}
		\label{fig:Sysmodel}
	\end{center} \vspace*{-0.4cm}
\end{figure*}

\section{System Model} \vspace{-0.1cm}
In this section, we introduce our channel modeling strategy for RIS-assisted indoor and outdoor wireless communication systems in sub-6 GHz bands. Proposed channel model for the RIS-empowered wireless network is established by considering the 3D channel modeling approach in 3GPP \cite{3GPP_Sub6G}. Additionally, the end-to-end single-input single-output (SISO) channel model is derived in the presence of an RIS. Furthermore, the near-field channel between the RIS-Rx is analyzed when the far-field conditions are not satisfied due to the close distance between the RIS-Rx. Since 2D channel models are not accurate for the RIS-assisted transmission owing to existence of an elevation angle, a 3D channel model should be taken into consideration. The considered propagation environments are Indoor Hotspot (InH) and Urban Microcellular (UMi) in this study. The generic system model for the RIS-assisted wireless communication system is illustrated in Fig. \ref{fig:Sysmodel}, where $d_\text{3D}^\text{T-RIS}$,$d_\text{3D}^\text{T-R}$ and $d_\text{3D}^\text{RIS-R}$ represents the 3D distances between the Tx-RIS, Tx-Rx and RIS-Rx, respectively. It is assumed that the RIS with $N$ number of elements is positioned on the $xz$-plane. There exist $C$ clusters in the environment, each containing $S$ rays. The channels between the Tx-RIS, RIS-Rx and Tx-Rx are denoted by $\mathbf{h}\in \mathbb{C}^{N\times 1}$, $\mathbf{g}\in \mathbb{C}^{N\times 1}$, and $h_\text{SISO}\in \mathbb{C}^{1\times 1}$, respectively, where $N$ is the number of RIS elements. It is assumed that the Tx and Rx are equipped with unity gain isotropic antennas. The positions of the Tx, Rx and RIS are given in the cartesian coordinate system as $\textbf{r}^\text{Tx}=(x^{\text{Tx}},y^{\text{Tx}},z^{\text{Tx}})$, $\textbf{r}^\text{Rx}=(x^{\text{Rx}},y^{\text{Rx}},z^{\text{Rx}})$ and $\textbf{r}^\text{RIS}=(x^{\text{RIS}},y^{\text{RIS}},z^{\text{RIS}})$, respectively. 

Considering the direct path and RIS reflection paths, the baseband equivalent of the received symbol at the Rx is expressed as \vspace*{-0.2cm}
\begin{align} \label{eq:received}
    y=\sqrt{P_t}(\textbf{g}^\mathrm{T}\mathbf{\Theta}\textbf{h}+h_\text{SISO})s+w \vspace{-0.2cm}
\end{align}
where $P_t$ is the transmit power, $s$ is the transmitted symbol, $w$ is additive white Gaussian noise at the Rx, which is modeled as $w\sim \mathcal{CN}(0,N_0)$, and $(\cdot)^{\mathrm{T}}$ denotes the transpose of the vector or matrix. The response matrix of the RIS is expressed as $\mathbf{\Theta}= \text{diag}(\beta_1e^{j\alpha_1}, \cdots, \beta_Ne^{j\alpha_N})$, where $\beta_n$ and $\alpha_n$ represents the magnitude and the phase shift of the $n$th RIS element for $n= 1,2,\dots,N$, respectively.

\subsection{Tx-RIS Channel}
\begin{figure}[!t]
\centerline{\includegraphics[width=0.65\columnwidth]{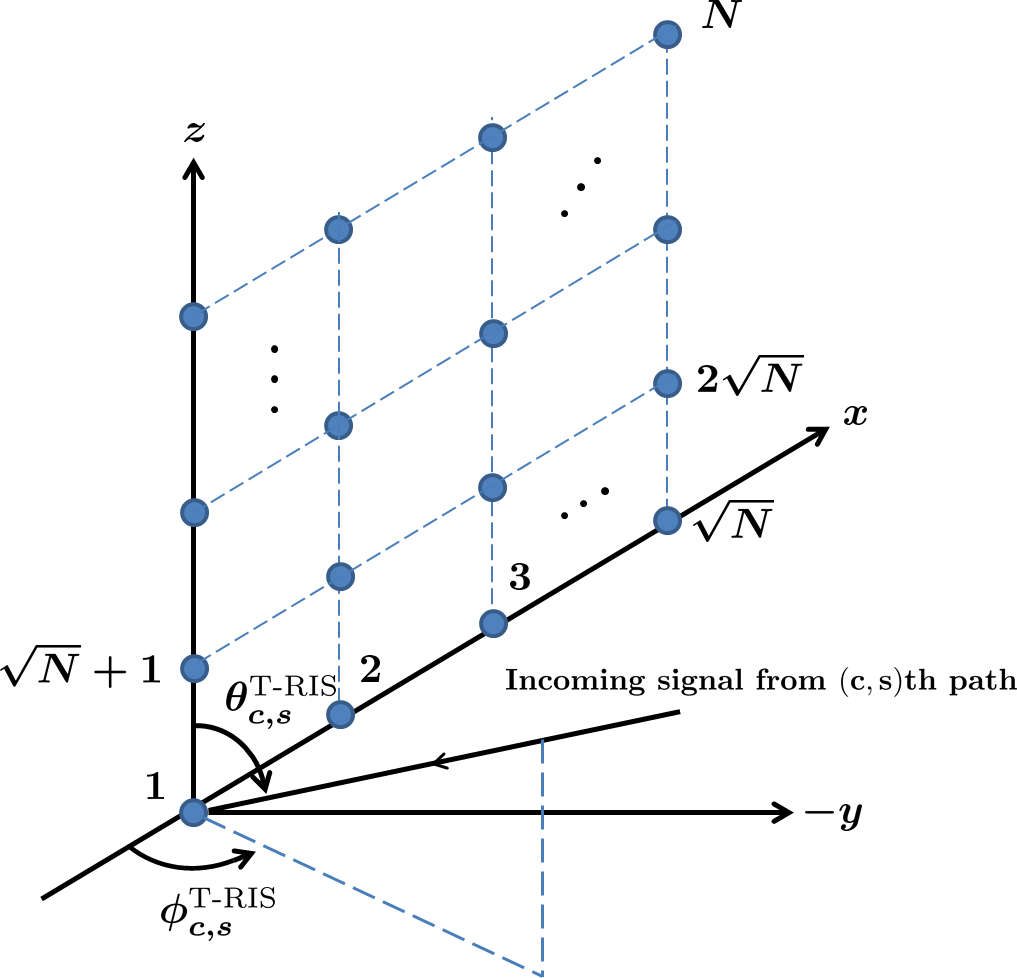}}
\caption{3D RIS geometry.} \vspace*{-0.4cm}
\label{fig:RIS_geometry} 
\end{figure}

In this subsection, the channel between the Tx and RIS is presented and the channel generation procedure is explained in detail. The channel $\mathbf{h}$ is expressed as \vspace{-0.1cm}
\begingroup\makeatletter\def\f@size{9}\check@mathfonts\begin{align}\label{eq:1}
    \mathbf{h}=\sum\limits_{c=1}^{C}\sum\limits_{s=1}^{S}\sqrt{\frac{P_c}{S}\vphantom{\frac{G_e(\theta_{c,s}^{\text{T-RIS}})}{P_L}}}\sqrt{\frac{G_e(\theta_{c,s}^{\text{T-RIS}})}{P_L}}e^{j\Phi_{c,s}}\mathbf{a}(\theta_{c,s}^{\text{T-RIS}},\phi_{c,s}^{\text{T-RIS}})
\end{align}\endgroup
where $P_c$, $G_e(\theta_{c,s}^{\text{T-RIS}})$ and $P_L$ are the power of the $n$th cluster, radiation pattern of an RIS element in the direction of $(c,s)$th path and the path loss component, respectively. Here, $\Phi_{c,s}$, $\theta_{c,s}^{\text{T-RIS}}$, and $\phi_{c,s}^{\text{T-RIS}}$ also represent the random initial phase, zenith angle of arrival (ZoA) and azimuth angle of arrival (AoA) of the RIS for the $(c,s)$th path, respectively. The array response vector of the RIS is also denoted by $\mathbf{a}(\theta_{c,s}^{\text{T-RIS}},\phi_{c,s}^{\text{T-RIS}})$. Considering these parameters, the generation steps of the Tx-RIS channel are listed as follows:
\begin{enumerate}[wide, labelwidth=!, labelindent=0pt]

    \item[\textit{1)}] First, we need to set the environment and the locations of the Tx and RIS. The line-of-sight (LOS) ZoA $(\theta_{LOS}^{\text{T-RIS}})$ and AoA $(\phi_{LOS}^{\text{T-RIS}})$ angles is calculated by using the locations of the Tx and RIS.
    
    \item[\textit{2)}] Then, we need to assign propagation conditions (LOS/NLOS) and calculate the path loss $P_L$. The LOS probabilities for InH and UMi scenarios are calculated as given in  \cite[Table 7.2.2]{3GPP_Sub6G}. In addition, we assume $100\%$ LOS probability, if the RIS is located at the same elevation or higher than the transmitter as $(z^{\text{RIS}} \geq z^{\text{Tx}})$. Reference path loss models \cite{3GPP_Sub6G} for LOS $(P_L^{\text{LOS}})$ and NLOS $(P_L^{\text{NLOS}})$ cases in the InH environment are given as follows, respectively:\vspace{-0.1cm}
    \begingroup\makeatletter\def\f@size{9.5}\check@mathfonts
    \begin{align} \label{eq:PL1}
        P_L^{\text{LOS}}\text{[dB]}=16.9\log_{10}(d_{\text{3D}})+32.8+20\log_{10}(f_c) + \sigma_{SF},\\ \label{eq:PL2}
        P_L^{\text{NLOS}}\text{[dB]}=43.3\log_{10}(d_{\text{3D}})+11.5+20\log_{10}(f_c) +\sigma_{SF}. \vspace*{-0.2cm}
    \end{align}
    \endgroup
    Reference path loss models \cite{3GPP_Sub6G} for LOS and NLOS cases in the outdoor UMi environment are given as follows, respectively:
    \begingroup\makeatletter\def\f@size{9.5}\check@mathfonts
    \begin{align} \label{eq:PL3}
        P_L^{\text{LOS}}\text{[dB]}=&22\log_{10}(d_{\text{3D}})+28+20\log_{10}(f_c)+\sigma_{SF},\\ \label{eq:PL4}
        P_L^{\text{NLOS}}\text{[dB]}=&36.7\log_{10}(d_{\text{3D}})+22.7+26\log_{10}(f_c)\nonumber\\&\hspace{2.85cm}\hspace{-0.74 cm}-0.3(h_{UT}-1.5)+\sigma_{SF}.
    \end{align}
    \endgroup
    In \eqref{eq:PL1}-\eqref{eq:PL4}, $f_c$ is the operating frequency in GHz, $\sigma_{SF}$ is the shadow fading term and $h_{UT}$ is the effective receive antenna height, which is the height of the RIS obtained by $h_{UT}=z^\text{RIS}-1$. Here, $d_\text{3D}$ is equivalent to $d_\text{3D}^\text{T-RIS}$ and calculated by $d_\text{3D}^\text{T-RIS}=\|\textbf{r}^\text{Tx}-\textbf{r}^\text{RIS}\|$.
    \item[\textit{3)}] Correlated large scale parameters should be generated.  Shadow fading (SF), Ricean $K$ factor, delay spread (DS), azimuth spread of arrival (ASA), azimuth spread of departure (ASD),  zenith spread of arrival (ZSA), and zenith spread of departure (ZSD) are generated by considering the cross correlation coefficients between each other given in \cite[Table 7.3-6]{3GPP_Sub6G}.
    \item[\textit{4)}] The delays and the cluster powers should be generated. The delay associated with the $c$th cluster is given as $\tau_c$ and randomly generated by using exponential power delay distribution \cite{3GPP_Sub6G}.
    The cluster powers depend on the cluster delays and follow an exponential power delay profile. The power of the $c$th cluster is denoted by $P_c$ and calculated as in \cite{3GPP_Sub6G}. Under LOS condition, the path from the first cluster is considered as the LOS path. Hence, the power of the LOS component $(P_{LOS}=K_R/(K_R+1))$ is added to the first cluster power and the power of the $c$th cluster is rewritten as \vspace*{-0.1cm}
   \begingroup\makeatletter\def\f@size{9}\check@mathfonts \begin{align}
        P_c=P_c/(K_R+1)+\delta(c-1)P_{LOS}\vspace*{-0.4cm}
    \end{align}\endgroup
    where $K_R$ is the Ricean factor in linear scale and $\delta(.)$ is the dirac delta function.

    \item[\textit{5)}] The ZoA $(\theta_{c,s}^{\text{T-RIS}})$ and AoA $(\phi_{c,s}^{\text{T-RIS}})$ angles of the RIS should be generated. Here, $\theta_{c,s}^{\text{T-RIS}}$ follows Laplacian distribution in both indoor and outdoor environments, and $\phi_{c,s}^{\text{T-RIS}}$ follows Laplacian distribution for indoors while it follows Wrapped Gaussian distribution for outdoors. The detailed generation process for $\theta_{c,s}^{\text{T-RIS}}$ and $\phi_{c,s}^{\text{T-RIS}}$ is explained in \cite[sec. 7.3]{3GPP_Sub6G}. Additionally, $\theta_{c,s}^{\text{T-RIS}}$ is distributed within the range of $[0^{\circ},180^{\circ}]$, while $\phi_{c,s}^{\text{T-RIS}}$ is distributed within the range of $[-180^{\circ},180^{\circ}]$. Here, we need to ignore clusters behind the RIS by limiting the range of $\phi_{c,s}^{\text{T-RIS}}$ to $[0^{\circ},180^{\circ}]$.
    
    \item[\textit{6)}] The initial random phase for the $(c,s)$th path should be generated, where $c$ is the cluster index and $s$ is the ray index in each cluster. Here, $\Phi_{c,s}$ follows an uniform distribution as $\Phi_{c,s}\sim\mathcal{U}(-\pi,\pi)$.
    
    \item[\textit{7)}] The radiation pattern of the RIS elements should be calculated. We consider the $\cos^q$ radiation pattern, which is used for reflectarray antennas \cite{nayeri2018reflectarray} to model the radiation of the RIS elements. The radiation pattern in the direction of the $(c,s)$th path is calculated as \vspace*{-0.1cm}
    \begin{align}
        G_e(\theta_{c,s}^{\text{T-RIS}})= 2(2q+1)\cos^{2q}(90-\theta_{c,s}^{\text{T-RIS}})\vspace*{-0.3cm}
    \end{align} 
    where $2(2q+1)$ is normalization factor and $q$ is obtained as $q=0.285$ in \cite{basar2020simris} by considering the physical area of RIS elements.
    
    \item[\textit{8)}] The array response vector of the RIS $\mathbf{a}(\theta_{c,s}^{\text{T-RIS}},\phi_{c,s}^{\text{T-RIS}}) \in \mathbb{C}^{N\times 1}$ should be calculated. Here, a square RIS structure is considered and the horizontal/vertical distances between the elements are equal and denoted by $d$. As illustrated in Fig.  \ref{fig:RIS_geometry}, the ZoA and AoA angles to the RIS are $\theta_{c,s}^{\text{T-RIS}}$ and $\phi_{c,s}^{\text{T-RIS}}$, respectively. Considering the RIS geometry in Fig. \ref{fig:RIS_geometry}, the first RIS element with the location vector $\textbf{r}^1=(x^1,y^1,z^1)$ is positioned at the bottom left corner of the RIS. $\textbf{r}^n=(x^n,y^1,z^n)$ denotes the location  vector of the $n$th RIS element, and $x^n$ and $y^n$ are respectively calculated as
    \begingroup\makeatletter\def\f@size{9}\check@mathfonts\begin{align}\label{eq:RIS_locs}
        x^n&=x^1+d\cdot\hspace*{-0.33 cm}\mod{\left(n-1,\sqrt{N}\right)},
      \nonumber  \\
        z^n&=z^1+d\cdot\bigg\lfloor (n-1)/{\sqrt{N}} \bigg\rfloor
    \end{align}\endgroup
    where $\hspace{-0.3 cm}\mod(.)$ is the modulus operation and $\lfloor . \rfloor$ is the floor operation.
    The normalized location of the $n$th RIS element according to first element is denoted by $(\overline{x}_n,0,\overline{z}_n)$, where $\overline{x}_n=x^n-x^1$ and $\overline{z}_n=z^n-z^1$.
     The array response vector for an RIS with $N$ elements is obtained by
    \begingroup\makeatletter\def\f@size{8}\check@mathfonts\begin{align}
    &\mathbf{a} (\theta_{c,s}^{\text{T-RIS}},\phi_{c,s}^{\text{T-RIS}} ) = \left[ 
    1 \; \cdots \; e^{j\frac{2\pi}{\lambda}( \overline{x}_n \cos \theta_{c,s}^{\text{T-RIS}} + \overline{z}_n\sin  \theta_{c,s}^{\text{T-RIS}} \cos\phi_{c,s}^{\text{T-RIS}}) }  \right. \nonumber \\&\cdots
    \left. e^{j\frac{2\pi}{\lambda} ( \overline{x}_N \cos \theta_{c,s}^{\text{T-RIS}} + \overline{z}_N\sin  \theta_{c,s}^{\text{T-RIS}} \cos \phi_{c,s}^{\text{T-RIS}}) } \right]^{\mathrm{T}}
    \end{align}\endgroup
    where $\lambda$ is the wavelength of the carrier signal for $n=1,\cdots,N$.
    
\end{enumerate}

Finally, we can form the channel $\mathbf{h}$ by using the channel parameters obtained in channel generation steps.

\subsection{Tx-Rx Channel}
The direct link between the Tx and Rx is given by
\begin{align}
    h_\text{SISO}=\sum\limits_{c=1}^{C}\sum\limits_{s=1}^{S}\sqrt{\frac{P_c}{S}}\sqrt{\frac{1}{P_L}}e^{j\Phi_{c,s}}
\end{align}
where $P_c$, $P_L$ and $\Phi_{c,s}$ is defined as in \eqref{eq:1}. $h_\text{SISO}$ can be obtained by following the Steps 1, 2, 3, 4 and 6 in the channel generation procedure described in Subsection II.A. In all steps, the Tx and Rx positions should be taken into account instead of Tx and RIS positions, respectively. Also, calculations for AoA and ZoA angles are not required in the generation process of $h_\text {SISO}$, since a SISO link established.
\subsection{RIS-Rx Channel}
For the RIS-Rx channel, we consider two different channel scenarios between the RIS and Rx. When the size of the RIS is large enough, the far-field distance between the RIS and Rx becomes significantly large. Therefore, the near-field channels are also considered in this section when the RIS is placed close to the Rx, while far-field channels are obtained by following the same procedure in previous sections.
\subsubsection{Far-Field Channel}
The far-field channel $\textbf{g}$ is given as \vspace{-0.2cm}
\begingroup\makeatletter\def\f@size{9}\check@mathfonts\begin{align} 
    \mathbf{g}=\sum\limits_{c=1}^{C}\sum\limits_{s=1}^{S}\sqrt{\frac{P_c}{S}\vphantom{\frac{G_e(\theta_{c,s}^{\text{T-RIS}})}{P_L}}}\sqrt{\frac{G_e(\theta_{c,s}^{\text{RIS-R}})}{P_L}}e^{j\Phi_{c,s}}\mathbf{a}(\theta_{c,s}^{\text{RIS-R}},\phi_{c,s}^{\text{RIS-R}})
\end{align}\endgroup
where $P_c$, $P_L$, and $\Phi_{c,s}$ is defined as in \eqref{eq:1} and $G_e(\theta_{c,s}^{\text{RIS-R}})$ is the RIS element radiation pattern in the direction of $(c,s)$th path. The azimuth of departure (AoD) and zenith of departure (ZoD) angles from the RIS are represented by $\phi_{c,s}^{\text{RIS-R}}$ and $\theta_{c,s}^{\text{RIS-R}}$, respectively. The array response vector of the RIS is denoted by $\mathbf{a}(\theta_{c,s}^{\text{RIS-R}},\phi_{c,s}^{\text{RIS-R}})$. The far-field channel can be produced by following the similar channel generation steps in Subsection II.A. In all steps, the RIS and Rx positions should be taken into account instead of Tx and RIS positions, respectively. In Steps 5, 7 and 8, the angles $\theta_{c,s}^{\text{RIS-R}}$ and $\phi_{c,s}^{\text{RIS-R}}$ should be considered instead of $\theta_{c,s}^{\text{T-RIS}}$ and $\phi_{c,s}^{\text{T-RIS}}$, and they follow the same distributions with $\theta_{c,s}^{\text{T-RIS}}$ and $\phi_{c,s}^{\text{T-RIS}}$, respectively.
Additionally, $\theta_{c,s}^{\text{RIS-R}}$ is distributed within the range of $[0^{\circ},180^{\circ}]$, while $\phi_{c,s}^{\text{RIS-R}}$ is distributed within the range of $[-180^{\circ},180^{\circ}]$. Here, we need to ignore clusters behind the RIS by limiting the range of $\phi_{c,s}^{\text{RIS-R}}$ to $[0^{\circ},180^{\circ}]$ as in Subsection II.A.

\subsubsection{Near-Field Channel}
In sub-6 GHz bands, the far-field distance, which is calculated by $\frac{N\lambda}{2}$ for a square RIS structure, increases significantly for larger RIS sizes. The scenarios, where the RIS is located in the far-field area, will be inefficient due to the high distance and multiplicative path loss. Hence, the near-field channels should be considered to analyze the RIS performance in near-field conditions. When the RIS operates in the near-field of Rx, there will be a pure LOS link between the RIS and Rx. Unlike far-field channels, the effective area of each RIS element changes since the angle of view of each RIS element to the Rx changes. Moreover, the distances from the RIS elements to the Rx varies. Furthermore, the polarization mismatch between the RIS elements should be taken into account under the near-field conditions\cite{bjornson2020power}. The near-field channel between the RIS and Rx is denoted by $\textbf{g}=[g_1,g_2,\cdots,g_N]^\mathrm{T}$, where $g_n$ represents the channel coefficient from the $n$th RIS element to the Rx, and can be expressed in terms of its magnitude and phase as $g_n=|g_n|e^{-j\gamma}$ for $n= 1,2,\dots,N$.
By considering the RIS geometry in Fig. \eqref{fig:RIS_geometry} and the RIS element locations in \eqref{eq:RIS_locs}, the near field channel gain from the $n$th RIS element is approximated as according to\cite{bjornson2020power}
\begingroup\makeatletter\def\f@size{9}\check@mathfonts\begin{align}\label{eq:nf}
    |g_n|^2\approx&\frac{1}{4\pi}\sum\limits_{x\in\mathbb{X}}\sum\limits_{z\in\mathbb{Z}} \left(\frac{\frac{xz}{y^2}}{3\left( \frac{z^2}{y^2}+1\right)\sqrt{\frac{x^2}{z^2}+\frac{z^2}{y^2}+1}}\right.\nonumber\\ 
    &\left.+\frac{2}{3}\tan^{-1}\left( \frac{\frac{xz}{y^2}}{\sqrt{\frac{x^2}{y^2}+\frac{z^2}{y^2}+1}}\right)\right)
\end{align}\endgroup
where $\mathbb{X}=\left\lbrace d/2+x^n-x^\text{Rx},d/2+x^\text{Rx}-x^n\right\rbrace$, $\mathbb{Z}=\left\lbrace d/2+z^n-z^\text{Rx},d/2+z^\text{Rx}-z^n\right\rbrace$, and $y=|y^n-y^\text{Rx}|$.
The phase of $g_n$ can be calculated as follows:
\begin{align}
    \gamma=2\pi\hspace{-0.3 cm}\mod{\left(\frac{\|\textbf{r}^n-\textbf{r}^\text{Rx}\|}{\lambda},1\right)}.
\end{align} \vspace*{-0.4cm}
\begin{figure}[!t]
	\begin{center}
		\includegraphics[width=0.9\columnwidth]{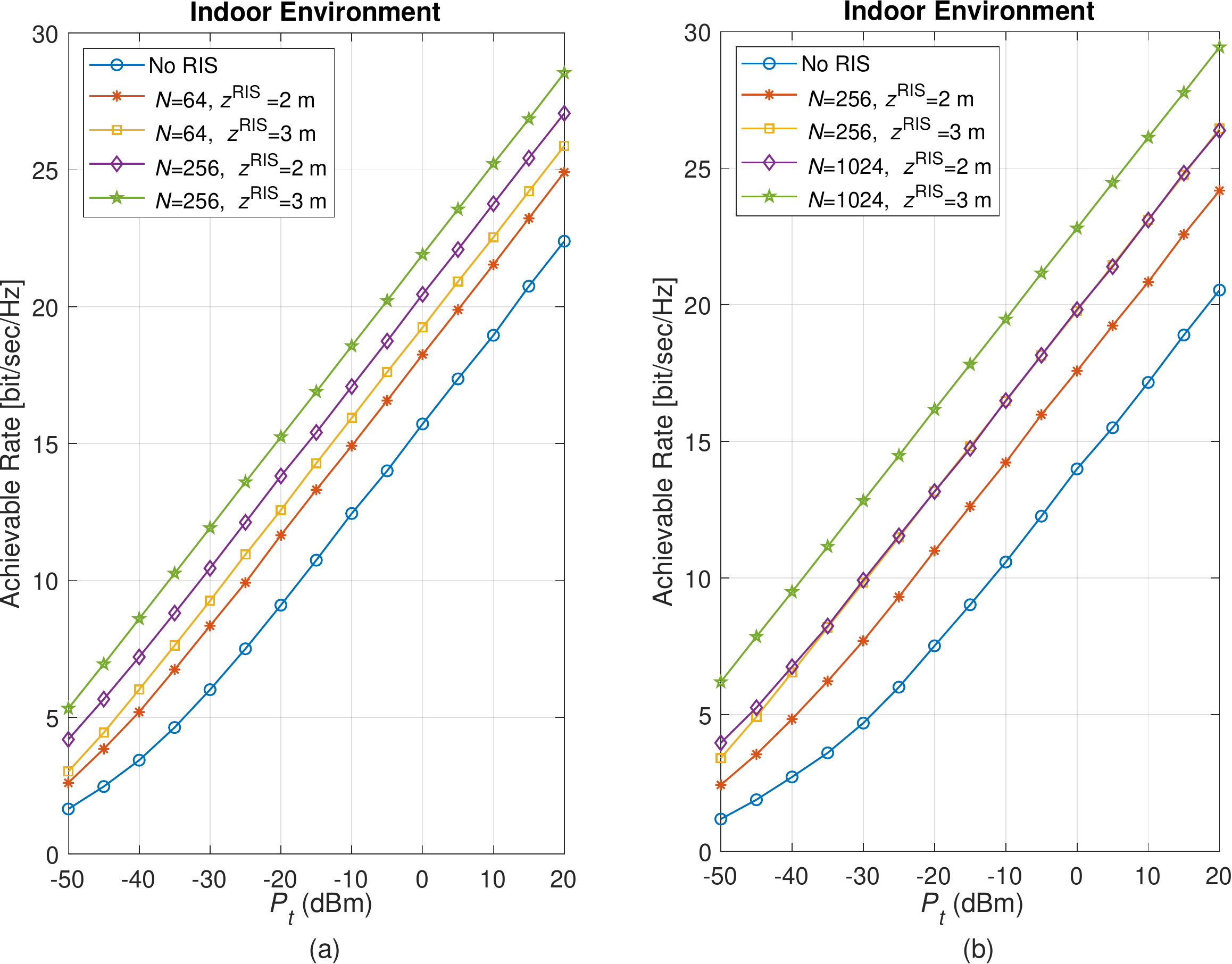}
		\caption{Achievable rates of RIS-assisted system for the indoor environment.}
		\label{fig:Sim1} \vspace*{-0.2cm}
	\end{center}\vspace*{-0.3cm}
\end{figure}
\section{Achievable Rate Analysis}
In this section, we analyze the achievable rate of the RIS-assisted communication systems by optimally adjusting the phases of the RIS elements according to the channel conditions. We assume that the magnitude response is $\beta_n=1$ for all RIS elements. The optimum phase response of the $n$th RIS element to maximize the signal-to-noise ratio (SNR) at the Rx is given by
\begin{align}\label{eq:RIS_phase}
    \alpha_n=\text{arg}(h_\text{SISO})-\text{arg}(h_ng_n)
\end{align}
where arg$(\cdot)$ denotes the phase of a complex number, $h_n$ and $g_n$ are the $n$th elements of $\textbf{h}$ and $\textbf{g}$, respectively.
By considering the phase response in \eqref{eq:RIS_phase} for the RIS elements, the instantaneous received SNR is expressed as
\begin{align}
    \rho=\frac{P_t|\textbf{g}^\mathrm{T}\mathbf{\Theta}\textbf{h}+h_\text{SISO}|^2}{N_0}=\frac{P_t\left|\sum\nolimits_{n=1}^N|h_n||g_n|+h_\text{SISO}\right|^2}{N_0}.\vspace*{-0.5cm}
\end{align}

The maximized achievable rate of the system can be computed by $R=\log_2(1+\rho)$.

\section{Simulation Results}
In this section, we provide computer simulation results for the RIS-assisted transmission in sub-6 GHz bands. The achievable rate of the RIS-assisted communication systems is investigated for different scenarios. The noise power $(N_0)$ is set to $-130$ dBm for all simulations. 

In Figs. \ref{fig:Sim1}(a) and (b), the achievable rates are presented for varying $N$ and different positions of the Tx, Rx and RIS for the indoor office environment. In both cases, the RIS is positioned in the near-field of the Rx, therefore, the near-field channels are considered for the RIS-Rx link under 2.4 GHz operating frequency. In Fig. \ref{fig:Sim1}(a), the considered 3D coordinates are $\textbf{r}^\text{Tx}=(0,25,3)$, $\textbf{r}^\text{Rx}=(40,48,1.5)$ and $\textbf{r}^\text{RIS}=(38,50,z^\text{RIS})$, while $\textbf{r}^\text{Tx}=(0,25,3)$, $\textbf{r}^\text{Rx}=(67,45,1.5)$ and $\textbf{r}^\text{RIS}=(70,50,z^\text{RIS})$ in Fig. \ref{fig:Sim1}(b). The achievable rate analysis is presented depending on the height of RIS in both scenarios. When the Tx and RIS are in the same elevation in case $z^{\text{RIS}}=3$ m, we assume $100\%$ LOS probability between the Tx and RIS. However, the LOS condition is not guaranteed in case $z^{\text{RIS}}=2$ since RIS is positioned below the Tx. According to Figs. \ref{fig:Sim1}(a) and (b), the achievable rate of the system is significantly better for the case $z^{\text{RIS}}=3$. Furthermore, the achievable rate of the system further increases as $N$ increases.

\begin{figure}[!t]
	\begin{center}
		\includegraphics[width=0.9\columnwidth]{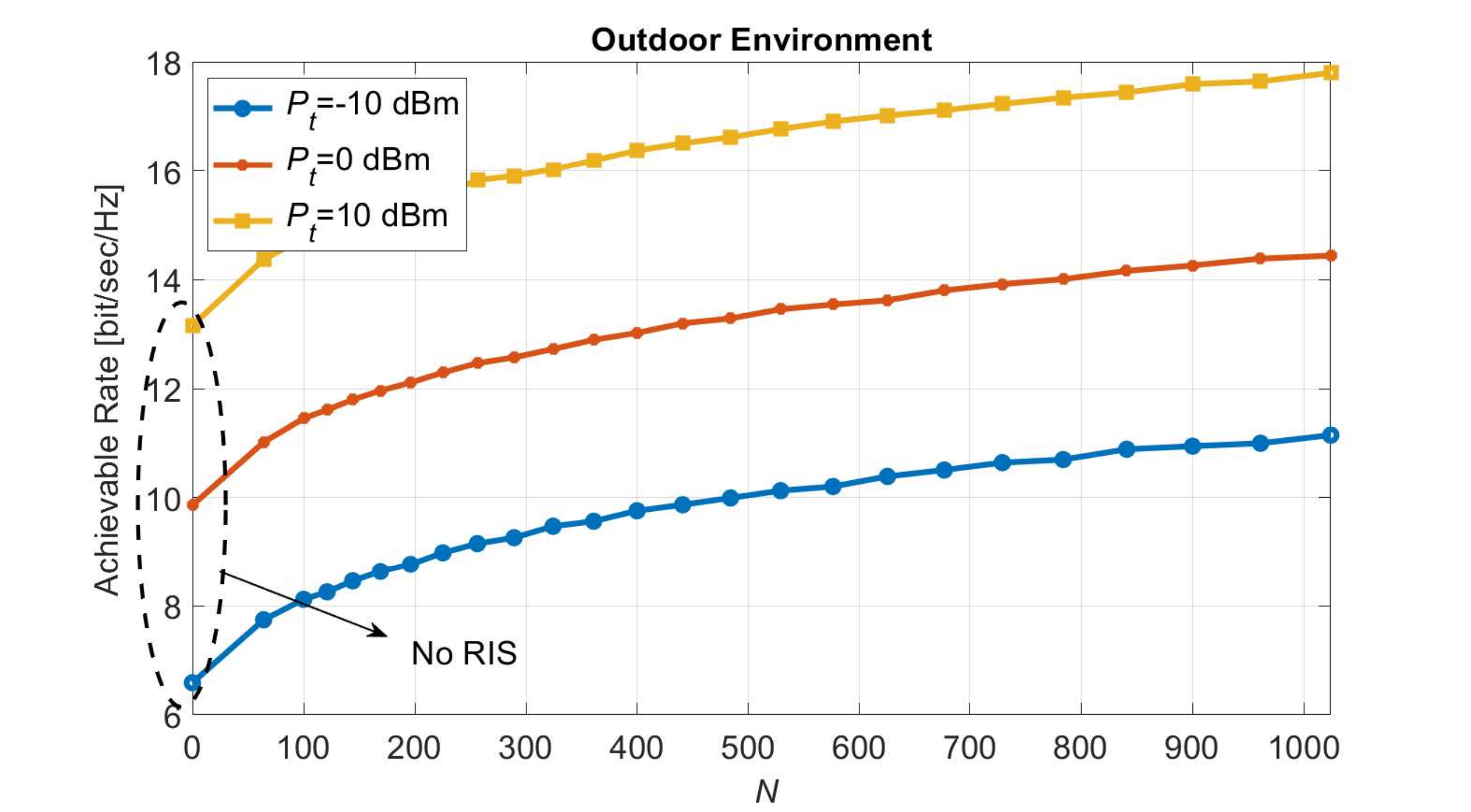}
		\caption{Achievable rates of RIS-assisted system for the outdoor environment.}
		\label{fig:Sim2}
	\end{center}\vspace*{-0.3cm}
\end{figure}

In Fig. \ref{fig:Sim2}, the effect of $N$ on the achievable rate is investigated in the outdoor environment. We consider 2.4 GHz operating frequency and the 3D coordinates as $\textbf{r}^\text{Tx}=(0,25,10)$, $\textbf{r}^\text{Rx}=(65,52,1)$ and $\textbf{r}^\text{RIS}=(62,55,7)$. Here, near-field channels are considered for the RIS-Rx link, since the RIS operates in the near-field of the Rx by increasing $N$. As seen from the Fig. \ref{fig:Sim2}, the improved achievable rate is obtained for increasing $N$ values. From the given results of Fig. \ref{fig:Sim2}, we observe that an RIS-assisted system with a sufficient number of RIS elements can outperform a system without RIS, which consumes 10 dB more transmit power.

\begin{figure}[!t]
	\begin{center}
		\includegraphics[width=0.95\columnwidth]{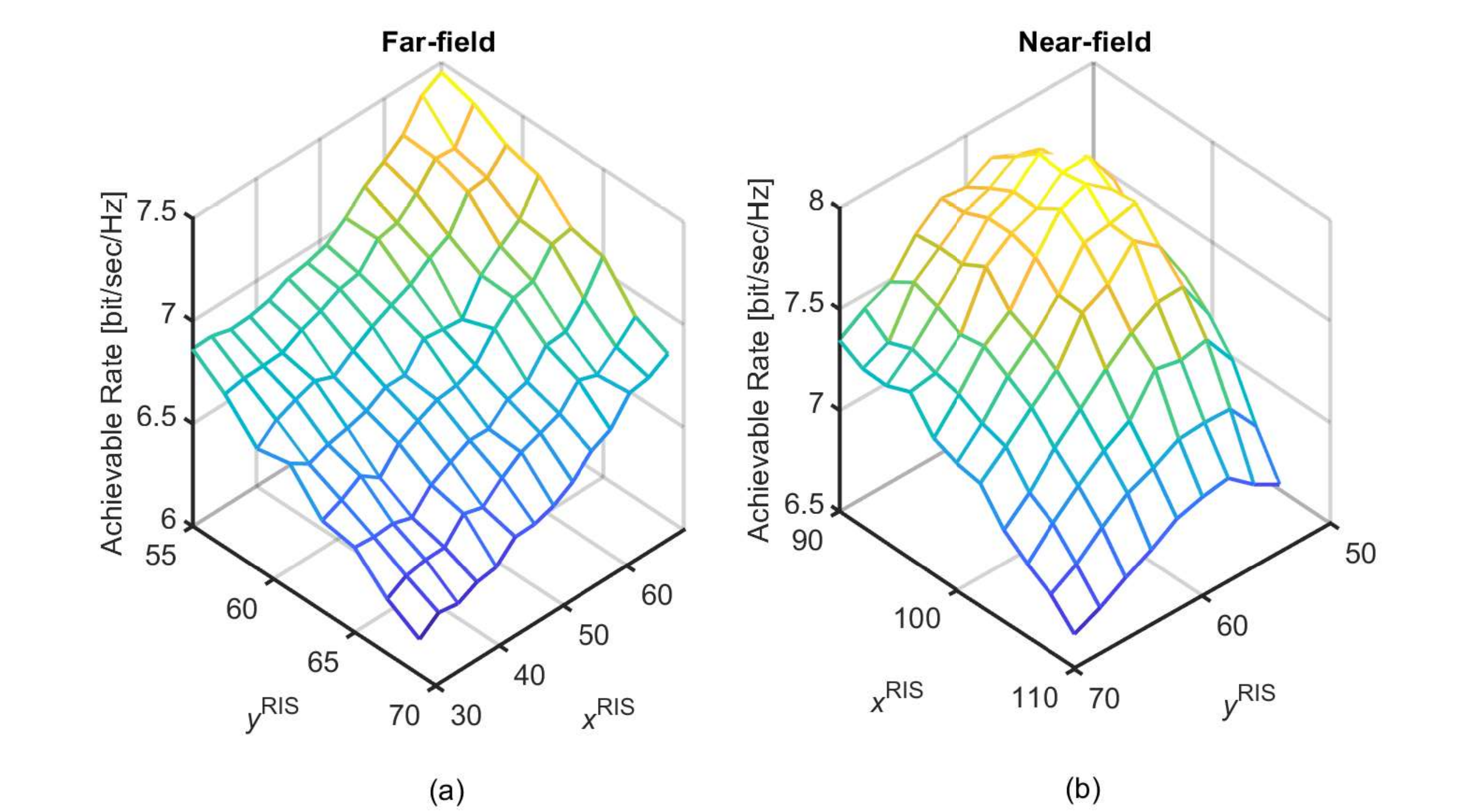}
		\caption{Achievable rates of RIS-assisted system for outdoors operating in (a) far-field and (b) near-field of Rx.}
		\label{fig:Sim3}
	\end{center}\vspace*{-0.6cm}
\end{figure}

In Figs. \ref{fig:Sim3}(a) and (b), the achievable rate of the system is presented under various RIS positions for the outdoor environment. We consider the operating frequency of 5.8 GHz, $N=1024$, and the 3D locations as $\textbf{r}^\text{Tx}=(0,25,10)$, $\textbf{r}^\text{Rx}=(100,50,1)$ and $\textbf{r}^\text{RIS}=(x^\text{RIS},y^\text{RIS},7)$. In Fig. \ref{fig:Sim3}(a), the RIS is operating in the far-field of the Rx, therefore, far-field channels are considered for the RIS-Rx link. As seen in Fig. \ref{fig:Sim3}, the achievable rate of the system increases as the RIS gets closer to the Rx in both $x$ and $y$-axes. Hence, the achievable rate of the system is maximized when $d_\text{3D}^\text{RIS-R}$ is minimized. When the RIS positioned in front of the Rx and operates in the near-field of the Rx, the achievable rate is maximized as observed in Fig. \ref{fig:Sim3}(b). Moreover, when the RIS is placed very close to the Rx around 2-3 meters in $y$-axis, the effective areas of the RIS elements decreases significantly due to the elevation difference between the RIS and Rx. Hence, the achievable rate is increasing for 2-3 meters as the RIS moves away in the $y$-axis as seen in Fig. \ref{fig:Sim3}(b). It should be noted that an improved performance in achievable rate is obtained when the near-field conditions are considered for large RISs.

\section{Conclusions}
In this paper, a physical channel model for RIS-assisted networks is presented in sub-6 GHz frequency bands. This study aims to meet the lack of analysis of RIS-assisted systems for sub-6 GHz bands in the literature by considering physical channels in sub-6 GHz bands. The performance of the system is analyzed from different perspectives, such as different environments, locations of the Tx, Rx and RIS, operating frequencies, number of RIS elements for near-field and far-field conditions. Our future works may include the extension of this work to MIMO systems along with real-world experimental results.\vspace*{-0.1cm}

\bibliographystyle{IEEEtran}
\bibliography{ref}

\end{document}